\documentclass[%
reprint,
amsmath,amssymb,
aip,
]{revtex4-1}

\usepackage{sidecap}
\usepackage{graphicx}
\usepackage{dcolumn}
\usepackage{bm}

\usepackage{float}

\begin{document}


\title{Broadband architecture for galvanically accessible superconducting microwave resonators }

\author{Sal J. Bosman}
\affiliation{%
Kavli Institute of NanoScience, Delft University of Technology,\\
PO Box 5046, 2600 GA, Delft, The Netherlands.
}%

  \author{Vibhor Singh}%
\affiliation{%
Kavli Institute of NanoScience, Delft University of Technology,\\
PO Box 5046, 2600 GA, Delft, The Netherlands.
}%

\author{Alessandro Bruno}
\affiliation{%
Kavli Institute of NanoScience, Delft University of Technology,\\
PO Box 5046, 2600 GA, Delft, The Netherlands.
}%

\affiliation{
Qutech Advanced Research Center, Delft University of Technology,
Lorentzweg 1, 2628 CJ Delft, The Netherlands.
}
\author{Gary A. Steele}

\affiliation{%
Kavli Institute of NanoScience, Delft University of Technology,\\
PO Box 5046, 2600 GA, Delft, The Netherlands.
}%




\date{\today}

\begin{abstract}
In many hybrid quantum systems, a superconducting circuit is required
that combines DC-control with a coplanar waveguide (CPW) microwave 
resonator.
The strategy thus far for applying a DC voltage or current bias to
microwave resonators has been to apply the bias through a symmetry
point in such a way that it appears as an open circuit for certain
frequencies.
Here, we introduce a microwave coupler for superconducting CPW
cavities in the form of a large shunt capacitance to ground.
Such a coupler acts as a broadband mirror for microwaves while
providing galvanic connection to the center conductor of the resonator.
We demonstrate this approach with a two-port $\lambda/4$-transmission
resonator with linewidths in the MHz regime ($Q\sim10^3$) that shows
no spurious resonances and apply a voltage bias up to $80$ V without
affecting the quality factor of the resonator.
This resonator coupling architecture, which is simple to engineer,
fabricate and analyse, could have many potential applications in
experiments involving superconducting hybrid circuits.

\end{abstract}

\maketitle

Embedding a quantum system, such as a qubit, in a microwave resonator
is an attractive and commonly used approach. Resonators provide
isolation that shields the system from environmental
noise and can controllably inhibit spontaneous decay. At millikelvin
temperatures, microwave resonators are in their ground state, free of
entropy, while still providing microwave frequency access to read-out
and manipulate quantum states. There is a wide range of hybrid systems 
exploring new quantum phenomena and technologies that require a 
microwave resonator that also offers DC-access to the device, 
including coupling mechanical resonators
to qubits \cite{lahaye2009nanomechanical, pirkkalainen2013hybrid},
microwave storage and conversion circuits \cite{andrews2015quantum},
Josephson and quantum dot radiation \cite{hofheinz2011bright,
  rokhinson2012fractional, chen2014realization, liu2015semiconductor},
spin qubits \cite{petersson2012circuit}, circuits coupled to ultra cold
atoms \cite{bothner2013inductively}, and more
\cite{xiang2013hybrid}. A cavity with galvanic access allows the
possibility to measure simultaneously a device's DC response, such as
a current-voltage curve, and its microwave response, like
emitted radiation or scattering characteristics. Such a setup could
form a bridge between DC quantum transport measurements and
all-microwave setups employed in circuit QED, with a wide range of
applications from the study of topological and other exotic
junctions\cite{moore2012viewpoint}, to superconducting molecular
junctions\cite{skoldberg2008spectrum}, to carbon
nanotubes\cite{ranjan2015clean,schneider2012coupling}.

\begin{figure}
\includegraphics[width=8.5cm]{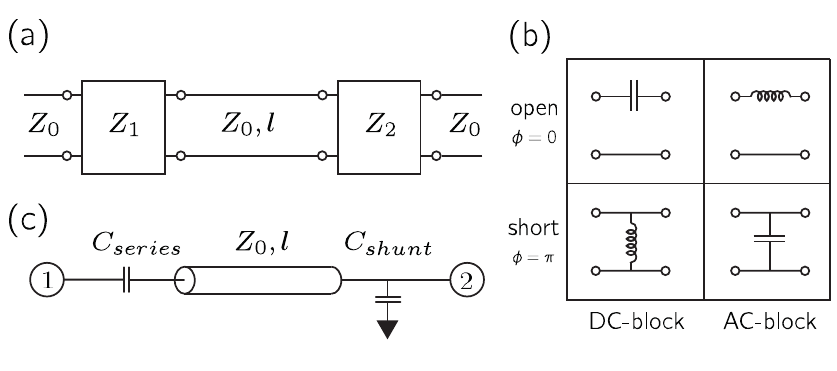}
\caption{(a) Schematic of a generic two-port cavity for confining
  electromagnetic fields using impedance mismatches as mirrors. The
  fields inside the cavity are isolated from the input-output fields
  by local impedance mismatches $Z_1$ and $Z_2$. (b) For a microwave
  CPW, mirrors can be implemented by incorporating
  capacitor or inductor elements either in series with the center
  conductor or from the center conductor to ground. This defines either an 
  quasi-open or -short boundary condition for microwaves and a conducting 
  or non-conducting path for DC signals. (c) Schematic of
  a cavity design implemented here providing DC access to the center
  conductor of the cavity using a shunt-capacitor as a mirror at port
  2. }
\end{figure}

\begin{figure*}
\includegraphics{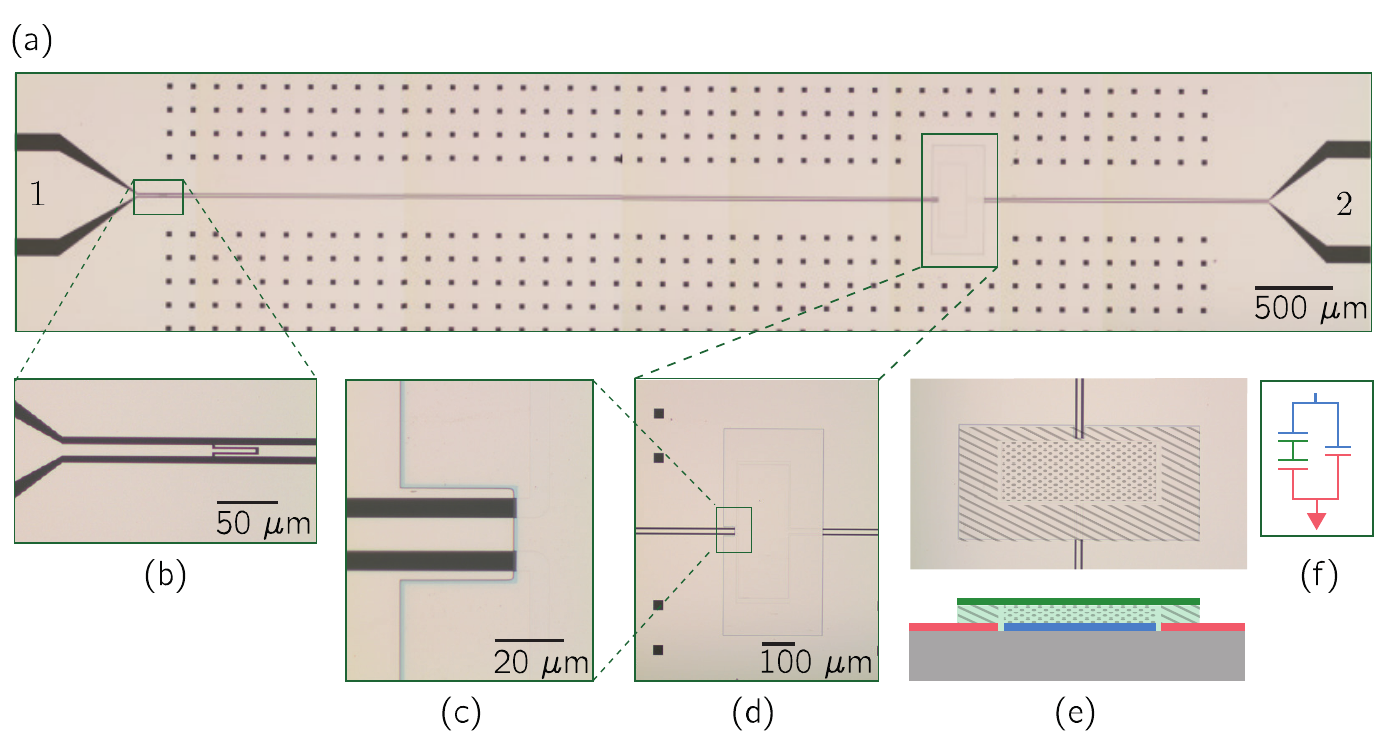}
\caption{Device: (a) Optical microscope image of the complete
  $\lambda/4$-transmission cavity. Port 1, on the left, is coupled
  with a gap capacitor, and on the right, port 2, is coupled through a
  shunt capacitor. (b) Shows a zoom in of the gap capacitor, (c) a zoom in on
  the shunt capacitor, with black (sapphire), blue
  ($\mbox{Si}_3\mbox{N}_4$), beige (MoRe). (d) Shows an overview of the shunt
  capacitor, (e) schematic top-view (above) and cross section (below)
  of the shunt capacitor. The center conductor is depicted in blue, the
  ground plane in red and the top-plate of the capacitive coupler in
  green. (f) Shows the equivalent circuit of the shunt capacitor. }
\end{figure*}

While attractive for many applications, applying a DC current or
voltage bias to a superconducting resonator without sacrificing its
quality factor is a non-trivial challenge \cite{chen2011introduction,
  li2013applying,de2014galvanically,hao2014development}.  The first
approach to incorporate DC-control into a superconducting microwave
resonator was to access the resonator
galvanically at a voltage node with a bias line made from a
$\lambda/2$-section of transmission line\cite{chen2011introduction}.  If the frequencies of
resonator and the bias line are perfectly matched, then the bias line
loads the circuit with an infinite impedance  (an effective open), resulting in
no leakage of the microwave field on resonance.  However, off-resonant
circuit excitations, such as a detuned qubit, are free to decay into
the bias-line.  This issue was addressed
recently\cite{hao2014development} using a reflective T-filter, making
the suppression band a few GHz wide and reaching quality factors $\sim
10^5$, at the cost of increased complexity. The insertion of the bias
into a symmetry point, typically breaking ground plane symmetry, makes
these designs susceptible to slot-line modes and spurious
resonances. Particularly for more complex circuits, these resonances can
complicate design, operation and analysis of the device.  In a third
approach\cite{de2014galvanically}, a lumped element resonator circuit
was split symmetrically in two such that a DC voltage can be applied
to the two halves using isolated ground planes without any radiation
losses, leading to high quality factors for perfectly balanced
designs.

Here, we explore a different approach in which we replace the typical
gap capacitor used as an input coupler in CPW cavities
with a large shunt capacitance to ground. Doing so, we achieve a
highly reflective microwave mirror in which the center conductor of the
waveguide is still galvanically connected to the input line, allowing
the application of a DC current or voltage. Such a design has several
advantages: in contrast to resonant filter designs, the reflectivity
is broadband up to the self-resonance frequency of the shunt
capacitor. As no extra port is required for the DC-signal, any energy
that leaks through the bias line can contribute to the measurement
signal.  Finally, this approach does not rely on any symmetry
considerations of the cavity, simplifying design and analysis.

Fig.\ 1 shows a schematic of a generic transmission line cavity,
consisting of a waveguide that is isolated from the input and output
ports by impedance mismatches. At the impedance mismatch points, the
propagating waves in the transmission line are reflected and a
standing wave forms at resonance.  Depending on the choice of 
impedance mismatch, one creates a boundary condition for the microwave field
corresponding to either a voltage node (short), or a current
node (open).  To couple the cavity to external circuitry one, or
both, of the boundary conditions are relaxed, which causes part of the
power to be transmitted.  In microwave cavities, a voltage node can be
implemented by a short to ground, while a current node by a gap
capacitor. These also have analogues in optical cavities: for optics,
a short circuit boundary can be implemented by a semi-silvered mirror,
while an open-circuit boundary can be implemented by a magnetic
mirror\cite{esfandyarpour2014metamaterial}.

In CPW microwave resonators, partially transmitting impedance
mismatches are typically implemented using lumped-element components
such as inductors or capacitors. In general, there are four types of
lossless couplers possible, which are depicted in Fig.\@ 1b.
Depending on the choice and configuration of the inductor or capacitor
one obtains unity reflection ($|S_{ij}|=\delta_{ij}$) in the limit
$\omega \mapsto 0$ (DC-block) or $\omega \mapsto \infty$
(AC-block). Including that reflection on a short causes a $\pi$-phase
shift, each quadrant of Fig.\@1b can be classified according to its scattering
behaviour as $S_{ij}(\omega)=\pm \delta_{ij}$, in the appropriate
limit \cite{bosman2015simple}.
\begin{table}[!b]
\begin{ruledtabular}
\begin{tabular}{llllll}
$|\Gamma|$ & $Q_c$ & $C_{series}$ [fF] & $L_{series}$ [nH] & $C_{shunt}$ [pF] & $L_{shunt}$ [pH]  \\
\hline
.5 & 3.14  & 318 & 3.18 & 1.27 & 796\\
.9 & 254 & 35.4 & 28.6 & 11.5 & 88.4\\
.99 & $31 \cdot 10^3$  & 3.22 & 315 & 126 & 8.03\\
.999 & $3.1 \cdot 10^6$ & .319 & 3180 & 1272& 0.79\\
\end{tabular}
\end{ruledtabular}
\caption{Typical values for different couplers; reflection
  coefficients, $|\Gamma|$, coupling $Q_c$'s, and the required value
  of capacitance or inductance, for a single port cavity of $5$ GHz of
  $50\ \Omega$ coupled with the coupler to a feedline of $50\ \Omega$.}
\end{table}

For a $\lambda/4$ CPW resonator made from a transmission line coupled
to a single external port, both of impedance $Z_0$, the coupling
quality factor $Q_c$ for all four types of couplers can be written in
the following unified form \cite{bosman2015simple}:
\begin{equation}
Q_c = \frac{\pi}{4} \frac{1}{b_c^{\pm 2}},
\end{equation}
where $b_c$ is the normalized susceptance of the coupler with $b_c=
Z_0 \omega C$ for capacitive couplers and $b_c=Z_0/L \omega$ for
inductive couplers, with $+/-$ for a series / shunt configuration
respectively. In Table 1 we tabulated the values of the inductor and
capacitor components required to achieve typical values $Q_c$ and
reflectivities $\Gamma$.

In choosing which microwave coupler is most suitable for galvanic
access, either the series inductor or the shunt capacitor,
consideration of stray reactance is important.  For example, a
resonator with $Q_c \sim 10^3$ coupled with a series inductor would
require an inductance of $100$ nH with a self-resonance frequency
above the cavity resonance. Realizing such a high on-chip inductance
with a self-resonance frequency above $\sim 8-10$ \ GHz is not
practical using geometric inductance, and requires high kinetic
inductance nano-inductors \cite{annunziata2010tunable} or Josephson
junction arrays \cite{castellanos2008amplification,
  manucharyan2009fluxonium}, which suffer from limiting dynamic range
and can complicate operation due to their non-linear nature.  In
contrast, using an on-chip parallel plate capacitor, as we will show
here, it is possible to engineer a large capacitance with small stray
inductance using conventional thin-film technologies.

As a proof of principle, we demonstrate the concept of a shunt
capacitor microwave mirror using a $\lambda/4$-transmission cavity
that incorporates one shunt capacitor and one series capacitor, shown
schematically in Fig.\@1c.  Fig.\@ 2 shows an optical microscope image
of the device using a CPW cavity and an on-chip
parallel plate capacitor.  The devices are fabricated in a three step
e-beam lithography process. First we pattern the resonator on sapphire
using a layer of $\sim 45$ \ nm of sputtered Molybdenum-Rhenium alloy with
reactive-ion etching\cite{singh2014molybdenum}.  Subsequently we
deposit $\sim 100$\ nm of PECVD $\mbox{Si}_3 \mbox{N}_4$, which is
wet-etched using HF. The upper plate of the capacitor is patterned in
a lift-off process using PMMA and a  MoRe film of $\sim 90$
nm. The resulting top plate acts as a capacitive coupler between the
center conductor and the ground plane, see Fig.\ 1e-f. We estimate the
parallel plate contribution to the shunt capacitance to be 27 pF and a stray 
capacitance from the lower capacitor electrode to the ground plane of $3-6$\ pF.
From formula 1, we estimate $Q_c=\pi Z_0^2 \omega^2 C^2/4\sim 2000$. 
Using finite element analysis simulations\cite{bosman2015simple},
we estimate the first self-resonance frequency of the capacitor to be $\sim 22$
\ GHz, significantly above the cavity frequency. From
equation 1, we need a gap capacitance at port 1 of $10$ fF to obtain a
symmetrically coupled cavity.

\begin{figure}
\includegraphics[width=8.5cm]{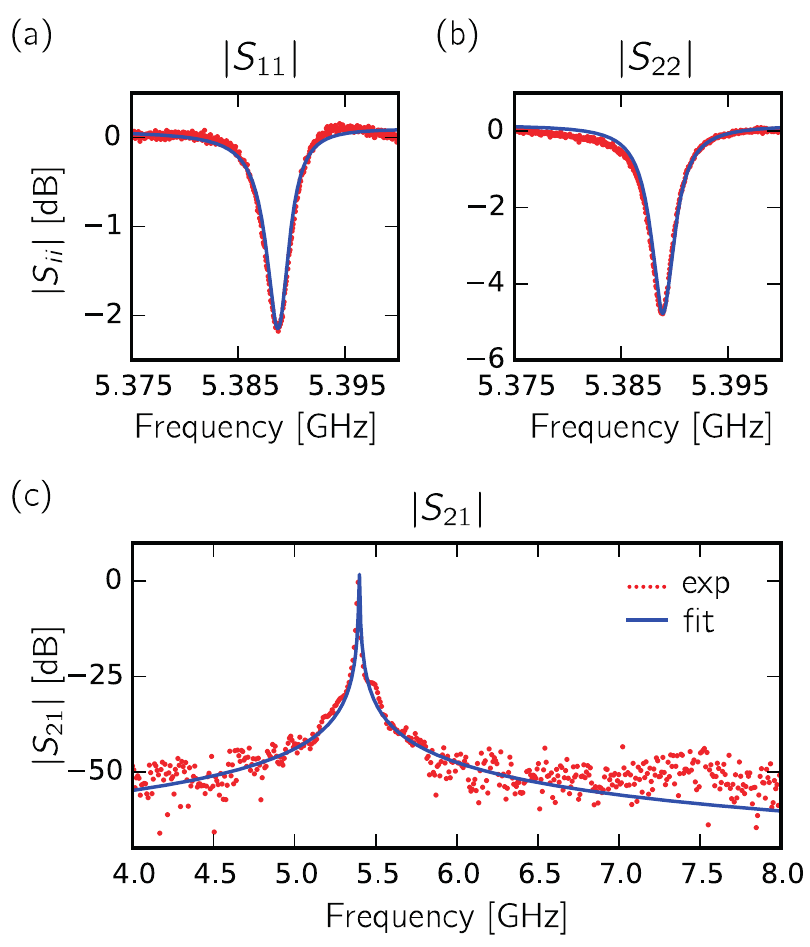}
\caption{Microwave network measurement characterizing the scattering
  matrix of the cavity measured with $-135$ dBm at the sample. (a)
  Reflection from port 1, measuring $\omega=2\pi \times 5.3889$ \ GHz,
  $\kappa_{tot}=2\pi \times 3.49$ MHz, and $\kappa_1=2\pi \times 0.51$
  MHz. (b) Reflection from port 2, measuring the same $\omega$, and
  $\kappa_{tot}$, with $\kappa_2=2\pi \times 2.75$ MHz. (c)
  Transmission from port 2 to 1, we measure a single resonance with
  the same frequency and linewidth.}
\end{figure}

For the microwave characterization, we cool the device to $T\sim 15
\ $mK in a radiation-tight microwave box. Using two identical
cryogenic microwave reflectometry configurations connected to both ports \cite{bosman2015simple}, 
we measure the full $S$-matrix of our device. We apply a DC voltage to the center conductor of
our cavity using a bias tee attached to port 2. Fig.\@ 3a shows a
measurement of the reflection from port 1 ($S_{11}$) with a resonance
at $\omega=2\pi \times 5.3889$ \ GHz, and a linewidth of
$\kappa_{tot}=2\pi \times 3.49$ \ MHz.  This demonstrates we are able to make
superconducting microwave resonators with loaded quality factors in the range
of $Q \sim 10^3$ that are galvanically accessible through a
frequency non-selective (non-resonant) connection.

In addition to measuring the cavity through the gap capacitance, we
can also use the shunt capacitor as a partially transmitting mirror.
Fig. 3b shows a measurement of the cavity resonance through the shunt
capacitor. We observe a resonance at the same frequency $\omega$ and linewidth, 
demonstrating that the coupler works as a mirror. 
From the fit \cite{bosman2015simple}, we find a coupling $Q_{c2}=1960$, giving an
effective capacitance of $29.5$ \ pF.  With the fits of both reflection measurements we can extract the coupling
rates $\kappa_{1,2}$ for each port \cite{bosman2015simple}. By combining these results 
we can extract the magnitude of the internal losses
using $\kappa_{tot} = \kappa_1 + \kappa_2 + \kappa_{int}$, and find an
internal loss rate of about $\kappa_{int} \sim 2\pi \times 230$\ kHz. 
We estimate\cite{bosman2015simple} a contribution of 8-80 kHz to the total 
internal loss rate of the cavity from the dielectric losses of the shunt capacitor, 
assuming a dielectric loss tangent $\delta \sim 10^{-3} - 10^4$. 
Reducing the dielectric thickness to 10 nm, we predict\cite{bosman2015simple} that this loss 
rate could be reduced to 0.8 kHz, limiting the internal Q of the cavity to
$6.3\times 10^5$.  

In transmission, Fig.\@3c, we observe a single resonance, with a
clean spectrum over the full measurement bandwidth of $4-8$ \ GHz,
showing that this approach does not suffer from spurious
resonances. 

\begin{figure}
\includegraphics[width=8.5cm]{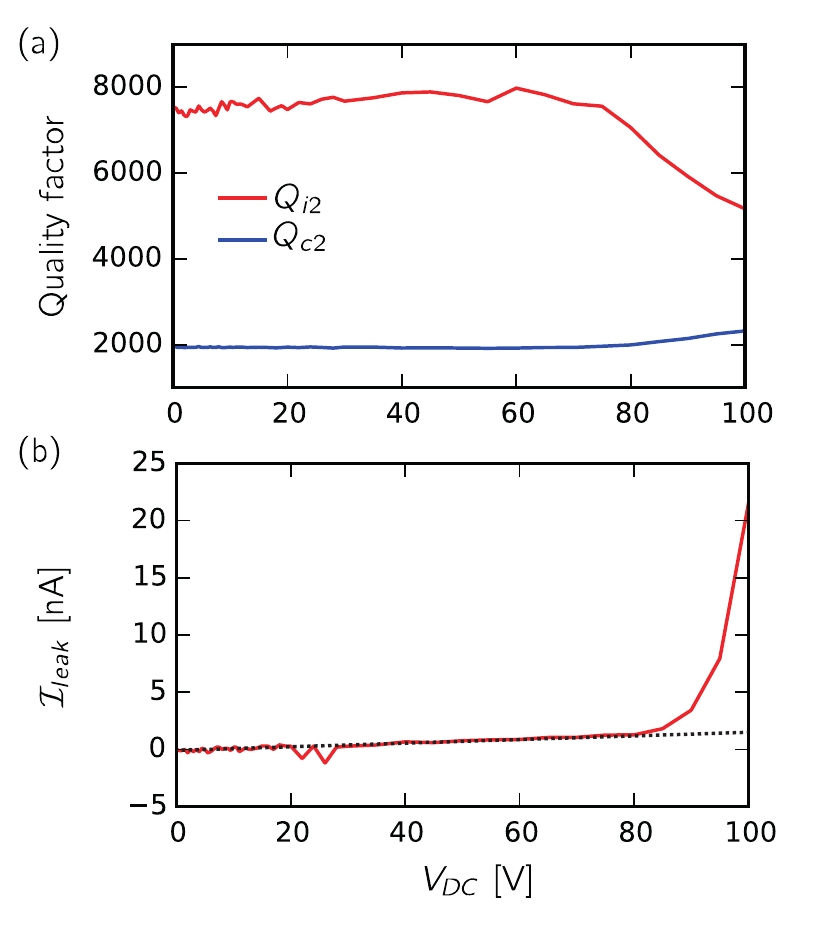}
\caption{Voltage bias characterization: (a) Coupling quality factor to
  port 2, $Q_{c2}$ and effective internal quality factor seen by port
  2, $Q_{i2}$, using a reflection measurement through the shunt
  capacitance ($S_{22}$), as a function of bias voltage measured at
  a power of $-135$ \ dBm, corresponding to $100$ photons. (b) Leakage
  current as a function of bias voltage. The dashed line indicates a
  resistance of $56\ \mbox{G}\Omega$.  }
\end{figure}

In figure 4, we characterize the performance of our cavity with a DC
voltage bias applied to port 2 using a bias tee. As shown in Fig.\@
4a, we observe no notable effect on the microwave response up to $\sim 80$
V. The same behaviour is also observed at single photon level (input
power of $-155$ dBm). At $80$ V, we observe a leakage current of $1.4$
\ nA, giving a resistance of $\sim 56\ \mbox{G}\Omega$. Above $80$ V,
we see the onset of dielectric break down with the leakage current
rising sharply to $25$ nA, leading to an estimated breakdown field of
$4$ MV/cm. Beyond 80 V, we see a degradation of the internal quality factor,
corresponding to an increase of $\sim 320$\ kHz to the internal loss rate. Also 
we observe a slight increase in the coupling quality factor to port 2, $Q_{c2}$, 
implying an increase in the shunt capacitance of about $\sim 5$\ pF. 
This could be caused by an inversion layer induced in the $\mbox{Si}_3\mbox{N}_4$ dielectric, 
effectively decreasing the plate separation. The increase of internal losses
could be related to Ohmic losses in the inversion layer.

To conclude, we have introduced a new type of microwave coupler that
allows galvanic access to the center conductor of a CPW microwave
resonator. We demonstrated this concept by engineering a
$\lambda/4$-transmission cavity with a high quality factor in which we
can apply a large DC voltage to the center conductor of the
resonator. By reducing the dielectric layer thickness to 10 nm, we predict 
that coupling quality factors of up to $3\times10^5$ should be possible 
while the first the self-resonance of the capacitor appears at $\sim 8$ GHz. \cite{bosman2015simple} 
For larger capacitances the dielectric losses of the deposited dielectric 
contribute less than a kHz to the internal loss rate of the cavity \cite{bosman2015simple}, 
at the expensive of a reduced dielectric breakdown voltage. 

The simplicity of this broadband technique
together with the possibility of large quality factors suggests that
this new design could be very attractive for many applications and
experiments involving superconducting hybrid circuits.

The authors would like to thank Daniel Bothner, Joshua Island, Nodar Samkharadze,
David van Woerkom, and Martijn Cohen for useful discussions.

\bibliography{bib_bosman}

\widetext
\clearpage
\begin{center}
  {\Large Supplementary Material: Broadband architecture for galvanically accessible superconducting microwave resonators \\ \ \\}
  Sal J. Bosman$^1$, Vibhor Singh$^1$, Alessandro Bruno$^{1,2}$, Gary A. Steele$^1$ \\ \ \\

  $^1${\em Kavli Institute of NanoScience, Delft University of Technology,\\
    PO Box 5046, 2600 GA, Delft, The Netherlands. \\ \ \\}
  $^2${\em Qutech Advanced Research Center, Delft University of Technology,
Lorentzweg 1, 2628 CJ Delft, The Netherlands.}
\end{center}


\renewcommand{\theequation}{S\arabic{equation}}
\renewcommand{\thefigure}{S\arabic{figure}}
\renewcommand{\thetable}{S\arabic{table}}

\renewcommand{\thesection}{S\arabic{section}}
\renewcommand{\thesubsection}{S\arabic{section}.\arabic{subsection}}
\renewcommand{\thesubsubsection}{S\arabic{subsection}.\arabic{subsubsection}}
\renewcommand{\bibnumfmt}[1]{[S#1]}
\renewcommand{\citenumfont}[1]{S#1}

\section{Measurement setup}

The complete measurement setup used for the device characterization is shown schematically in Fig.\@ S1. It consists of two identical microwave reflectometry measurement  chains, such that we have full access to the scattering matrix of the device ($S_{11},S_{12},S_{22},S_{21}$). The vector network analyser (VNA) outputs a signal that is fed through a variable attenuator at room temperature into the fridge, where it is first heavily attenuated before reaching the sample through a directional coupler. The reflected signal from the device is sent back to the VNA using two isolators and amplifiers. From each reflectometry setup we can measure the reflection of that port and by combining them we can measure transmission in both directions. Additionally we can apply a DC voltage to port 2 of the device through an unfiltered DC-line using a bias-tee just before the device.

\begin{figure}[b!]
\includegraphics[width=8cm]{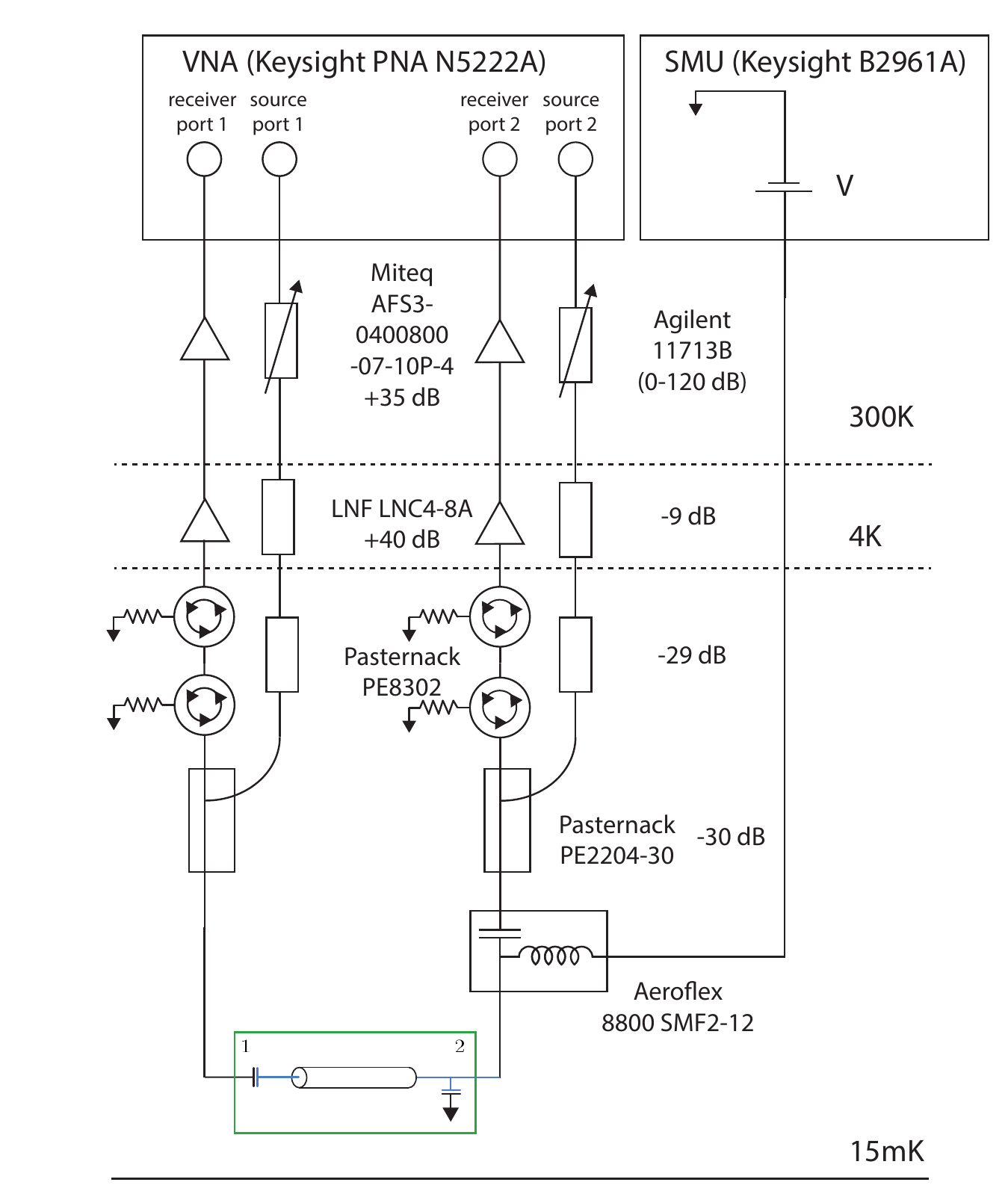}
\caption{\label{fig:wide} Measurement setup: In the dilution refrigerator we have two identical microwave reflectometry setups. The directional coupler near the sample separates the incident and reflected power from the sample. Additionally, the  measurement chain coupled to port 2 of the device has a bias-tee to inject the DC voltage bias onto the chip.  
 }
\end{figure}

\pagebreak

\section{\label{sec:level1}Microwave circuit analysis}

For two-port components, as depicted in Fig.\@1b of the main text, the transmission matrices, or $ABCD$-matrices are of the form [S1]:

\begin{equation}
T_{series}= \left( \begin{array}{cc} 1 & Z \\ 0 & 1  \end{array} \right), \hspace{40pt} T_{shunt}=\left( \begin{array}{cc} 1 & 0 \\ \frac{1}{Z} & 1 \end{array} \right), \hspace{40pt} \left( \begin{array}{c}V_1 \\ I_1\end{array} \right) = T \left( \begin{array}{c}V_2 \\ I_2\end{array} \right).
\end{equation}

These matrices relate incoming voltages and currents of port 1 ($V_1, I_1$) to outgoing ($V_2, I_2$) from port 2, hence characterize transmission of the two-port network. From these matrices we can already understand the limiting behaviour $S_{ij}(\omega)=\pm \delta_{ij}$ stated in the main text. If one takes the limit wherein the off-diagonal element goes to zero, the component shows perfect transmission. And in the limit where the off-diagonal element diverges, all power is reflected. The minus sign for reflection of a shunt component is due to the reversed sign of $I_2$ in $T$-matrices compared to scattering matrices (i.e. the current flows out of port  2). Using standard formulas we can rewrite $T$ into explicit scattering parameters for each component as:

\begin{equation}
S^{series}_{ii} = \frac{z}{z+2}, \hspace{30pt} S^{series}_{ij} = \frac{2}{2+z}, \hspace{30pt} S^{shunt}_{ii}=\frac{-1}{2z+1}, \hspace{30pt} S^{shunt}_{ij}=\frac{2z}{2z+1}.
\end{equation}

Using these expressions the reflection coefficient of port i, $\Gamma^{i}=S_{ii}$, can easily be calculated, where $z$ is the normalized impedance $z=Z/Z_0$ and $z=1/j Z_0 \omega C$  ($z=j \omega L/Z_0$) for the capacitor (inductor), with $j$ the imaginary unit. It is perhaps interesting to note that the $50-50$ beam splitter condition ($|\Gamma|=0.5$), for such a two-port component embedded in a $50\ \Omega$ transmission line (TL) needs a $25 \ \Omega$ impedance for a shunt and $100 \ \Omega$ for a series component 



To derive formula 1 in the main text we need to determine the coupling quality factor for each of the four coupler types  for a single-port $\lambda/4$-microwave resonator. A $\lambda/4$ resonator requires opposite boundary conditions on each end of the TL, hence the open type of couplers are connected to a shorted TL and vice versa for the short type of couplers.

Following reference [S1], here we present the generalized case. First we consider the open types of terminations; the gap capacitor and series inductor. In this configuration we have the coupler in series with a shorted TL, which has an impedance of $z_{TL}(\omega)=j \ \mbox{tan}(\beta l)$,  with $l$ the length of the TL and $\beta=\omega/v_p$, where $v_p$ is the phase velocity and $j$ the imaginary unit. At resonance we have $\beta l = \pi/2$, and the input impedance diverges. Now the impedance of the coupled resonator seen from the feedline is $z_{in}=z_c+z_{TL}$, with $z_c$ the impedance of the coupler, either $1/j Z_0 \omega C$ or $j \omega L/Z_0$. To perform a Taylor expansion around the resonance, where $z_{TL}$ diverges it is convenient to rewrite it in terms of the susceptance of the coupler $b_c=\mbox{Im}(1/z_c)$ and the cotangent. Note that for capacitors $z_c=-j/b_c$ and for inductors $z_c=j/b_c$, which we abbreviate as $\pm$ (upper sign corresponds to an inductive element and the lower sign for a capacitive element). Now we can write the input impedance as follows:

\begin{eqnarray*}
z_{in}(\omega) &=& j \bigg(\pm \frac{1}{b_c} + \frac{1}{\mbox{cot}(\beta l)}\bigg),\\
&=& j \bigg( \frac{b_c \mp \mbox{cot}(\beta l)}{b_c  \mbox{cot}(\beta l)}\bigg), \hspace{60pt} \mbox{at resonance: } b_c\pm \mbox{cot}(\beta l)=0, \ \ \mbox{(I)} \\
& \simeq & \pm j  \frac{\pi}{2} \frac{1+\mbox{cot}^2(\beta l)}{\mbox{cot}^2(\beta l)}\frac{(\omega - \omega_0)}{\omega_0}, \hspace{17pt} \mbox{where we Taylor expanded around \ } \beta l =\frac{\omega l}{v_p} \simeq \pi/2, \\
 & \simeq & \pm j  \frac{\pi}{2} \frac{1 + b_c^2}{b_c^2} \frac{(\omega - \omega_0)}{\omega_0}, \hspace{45pt} \mbox{where we used (I),} \\
 & \simeq & \pm j  \frac{\pi}{2} \frac{1}{b_c^2}\frac{(\omega - \omega_0)}{\omega_0}, \hspace{63pt} \mbox{if}\ b_c\ll 1. 
\end{eqnarray*}
The last approximation breaks down for very large coupling susceptances. Remark that we used the trigonometric
identity $1+\cot^2 x=1/\sin^2 x$ to rewrite $dz/d(\beta l)=\pm j/\cos^2(\beta l)=\pm j/(\cot^2(\beta l) \sin^2(\beta l))$. In this way we could invoke the resonance condition. For a low loss resonator we can include internal losses by substitution of $\omega_0 \rightarrow \omega_0(1\mp j/(2Q_{int}))$ in the numerator and obtain:

\begin{equation}
z_{in}(\omega)=\frac{\pi}{4Q_{int} b_c^2} \pm j\frac{\pi}{2}\frac{\omega-\omega_0}{\omega_0 b_c^2}.
\end{equation}
This expression of the input impedance is equivalent to that of a series LC-resonator, with a resistance on resonance of $R=Z_0\pi/(4Q_{int} b_c^2)$. At critical coupling the resonator on resonance is impedance matched to the feedline, and $Q_{int}$ equals $Q_c$, which is captured with the coupling factor $g=Z_0/R=Q_{int}/Q_c$. Now using $Q_c=Q_{int} R/Z_0$ we obtain the result:

\begin{equation}
Q_c = \frac{\pi}{4}\frac{1}{b_c^2}. 
\end{equation}
For the shunt couplers the derivation is very similar. Now we consider a shunt inductor ($\mbox{Im}(z_c) = 1/b_c=\omega L/Z_0$) or a shunt capacitor ($\mbox{Im}(z_c)=1/b_c=1/(Z_0 \omega C$) coupled to an open ended $\lambda/4$ TL, with $z_{TL}=-j \cot(\beta l)$. Now we have a parallel circuit so we obtain:
 
\begin{equation}
\frac{1}{z_{in}(\omega)}=j(\mp b_c +\frac{1}{ \cot(\beta l)}).
\end{equation}
We see that if we replace $b_c \rightarrow 1/b_c$ and reverse the sign we obtain the same expression, though in admittance $y_{in}=1/z_{in}$. This shows that such coupled cavities behave as parallel LC-resonators. Following the same logic we obtain:
\begin{equation}
y_{in}(\omega)=\frac{\pi b_c^2}{4Q_{int}} \mp j\frac{\pi}{2}\frac{\omega-\omega_0}{\omega_0 b_c^2},
\hspace{20pt} 
\end{equation}
with (+) for capacitors and (-) for inductors respectively. So we conclude that for the shunt couplers we obtain:

\begin{equation}
Q_c = \frac{\pi}{4}b_c^2.
\end{equation}
In the derivation for the series couplers we used $b_c \ll 1$, since for example a 10 fF gap capacitor in a 5 GHz cavity of 50 Ohms is $b_c \sim 0.02$, making the approximation valid. In the case of the shunt capacitor the reasoning goes exactly in opposite direction, where $b_c^{-1} \ll 1$, in the case of 30 pF for the same cavity we obtain $b_c \sim 50$, justifying this approximation.



\section{Microwave response}

With the expression for the input impedance at hand, it is easy to derive the microwave response of the cavity in reflection. Each coupling port contributes to the total loss rate $\kappa_{tot}$ and causes a small shift of the resonance frequency, which is here not of our interest.  In our implemented microwave cavity, as depicted in Fig.\@ 1c and Fig.\@ 2 of the main text, we can distinguish three separate loss channels that contribute to the total linewidth of the cavity: $\kappa_{tot}=\kappa_1 + \kappa_2 + \kappa_{int}$. A reflection measurement on port $i$ is sensitive to the ratio between the total linewidth and coupling rate to port $i$, captured with the coupling coefficient: $\eta_i = \kappa_i/\kappa_{tot}$. Therefore if we absorb the coupling rate of the other port into $\kappa_{int}$ we can suffice using the previously derived expressions. The reflection of the effective single-port cavity can then be written as:


\begin{equation}
S_{11}(\omega) = \frac{z_{in}-1}{z_{in}+1} = 1-\frac{2}{1+z_{in}}. 
\end{equation}
Using that $\omega_0/\kappa_1 = Q_{c1}$ and similar for $\kappa_{int}$ we can rewrite $z(\omega)$ (formula S3) as follows:

\begin{equation}
z_{in}(\omega) = \frac{\kappa_{int}}{\kappa_1}  - \frac{2j(\omega-\omega_0)}{\kappa_1}.
\end{equation} 
Combining both expressions and rearranging the $\kappa$'s we obtain:

\begin{equation}
S_{11}(\omega) = 1 - \frac{2 \eta_1}{1-2j(\omega-\omega_0)/\kappa_{tot}}
\end{equation}

The reflection response on port 2 ($S_{22}$) is very similar, with the only difference that it was written in admittance. Here we will use $z_{in}$ as the input impedance for port 2:

\begin{equation}
S_{22}(\omega) = \frac{z_{in}-1}{z_{in}+1}=\frac{y_{in}^{-1}-1}{y_{in}^{-1}+1}=\frac{1-y_{in}}{1+y_{in}}=-1+\frac{2}{1+y_{in}}.
\end{equation}
Following the same logic as before we obtain:

\begin{equation}
S_{22}(\omega) = -1 + \frac{2 \eta_2}{1+2j (\omega-\omega_0)/\kappa_{tot}}.
\end{equation}
For deriving the transmission response $S_{21}=S_{12}$ one cannot use this trick and the cavity should be analysed using transfer matrices. By multiplying the $T$-matrices of the gap-coupler, the TL and shunt coupler, one arrives at the $T$-matrix for the full cavity, from which an expression for $S_{12}$ can be determined. If we abbreviate $b_1=Z_0 \omega C_{gap}$ and $b_2=Z_0 \omega C_{shunt}$, and follow a similar procedure as before we obtain:

\begin{equation}
S_{12}(\omega) = \frac{2 \frac{b_1}{b_2}}{j(b_1^2 + b_2^{-2})+2(\omega-\omega_0)/(4\omega_0/\pi)}.
\end{equation} \\
This expression can then be simplified to:
\begin{equation}
S_{12}(\omega) = S_{21}(\omega) = \frac{2 \sqrt{\eta_1 \eta_2}}{j+2(\omega-\omega_0)/\kappa_{tot}},
\end{equation}
and we have obtained all response functions. For reflection of the gap capacitor ($S_{11}$, formula S10) with a large detuning from the resonance frequency, the response is anchored at the point $(1,0)$ in the complex plane, which shows that the reflection on a (quasi)-open does not cause a phase shift of the signal. The response of the shunt capacitor, described by formula S12, is anchored at $(-1,0)$, showing the $\pi$ phase shift on a (quasi)-short boundary condition. Transmission, $S_{21}=S_{12}$, for a large detuning is anchored at the origin, indicating there is no transmission of very off-resonant signals.

Close to resonance the term $2j(\omega-\omega_0)$ becomes important and the response will trace a (resonance) circle in the complex plane, where the radius depends on the coupling coefficient $\eta_i$. For an over-coupled port ($\eta_i>0.5$) the resonance circle will enclose the origin and the phase angle will change by $2\pi$, whereas the phase angle of an under-coupled port  will remain on the same branch. We would like to make a cautionary note on interpretting phase data without inquiring the resonance circle, as the `unwrapping' of different phase branches is arbitrary and can lead to erroneous conclusions. At resonance the magnitude of the reflection response will show a minimum (dip) of magnitude $|S_{ii}(\omega_0)|=|1-2\eta_i|$. After determining whether the port is under- or over-coupled, $\eta_i$ can be found by measuring the depth of the resonance dip. To illustrate the differences in the microwave response we show the results of a simulation in QUCs (an open-source circuit simulator [S2]) in Fig.\@ S2.

\begin{figure}
\includegraphics[width=16cm]{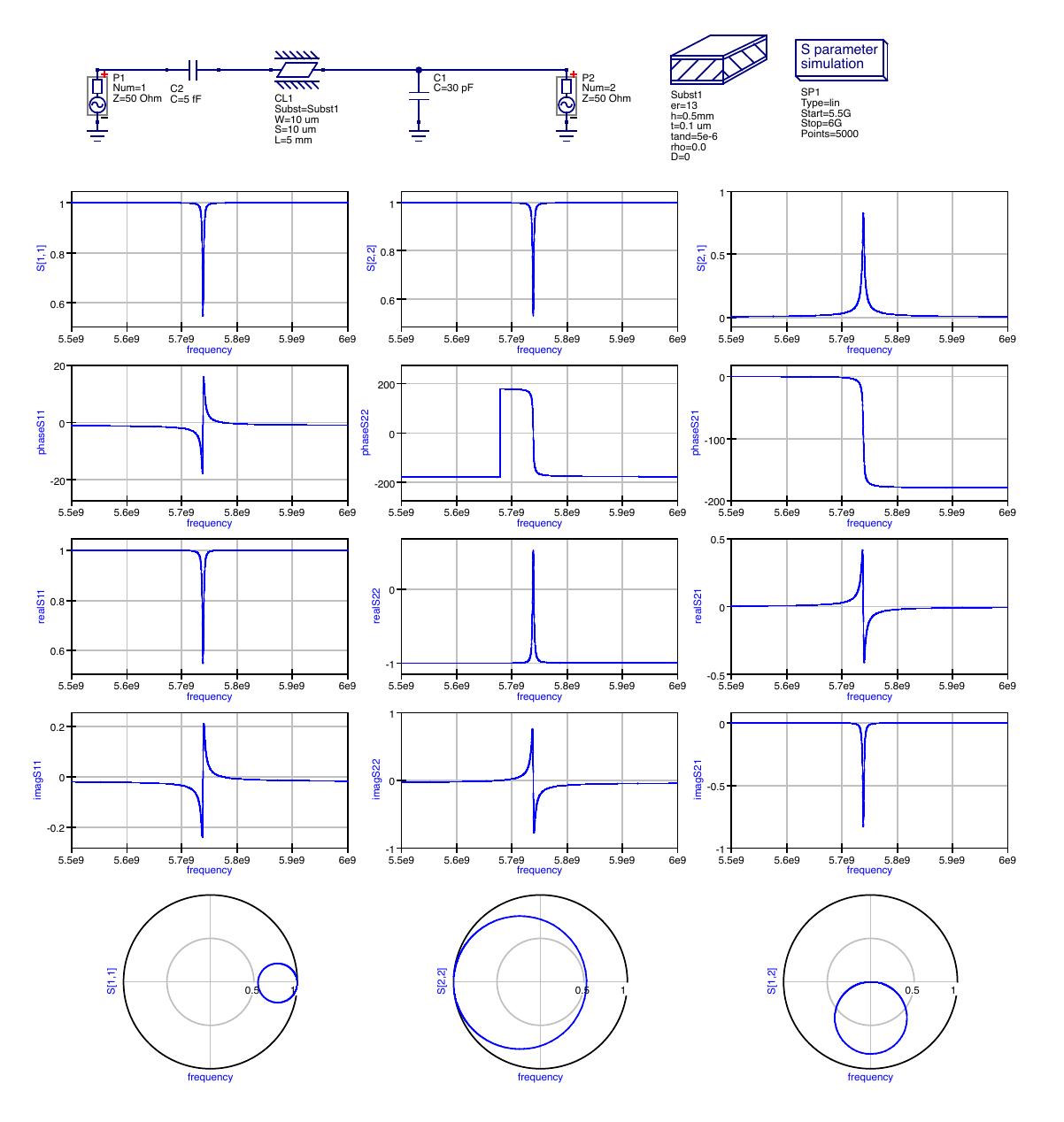}
\caption{\label{fig:wide}Numeric S-parameter simulation in QUCs of a cavity with a gap capacitance of 5 fF and a shunt capacitance of 30 pF (not the same values as in the experiment) illustrating the differences in microwave response; (left $S_{11}$ under coupled and anchored at (1,0), middle $S_{22}$ over coupled and anchored at (-1,0), and right shows $S_{12}$ anchored in (0,0) and its magnitude is equal to $\sqrt{\eta_1 \eta_2}$ and hence is sensitive to the coupling imbalance $\kappa_1/\kappa_2$ as well as to the magnitude of internal losses $\kappa_{int}$ (which is sometimes omitted).} 
\end{figure}

\newpage

\section{Data analysis}

Here we demonstrate the data analysis with the reflection data of port 2. Experimental realities like cable resonances and other uncontrolled factors in the circuitry can change the response from a Lorentzian line shape to a skewed-Lorentzian or Fano-line shape. This can be captured by the following adaptation of formula S12:

\begin{equation}
S_{22}(\omega) = \alpha e^{i\phi} + (1-\alpha)\bigg(-1 + \frac{2\eta_2}{1+2j (\omega-\omega_0)/\kappa_{tot})}\bigg),
\end{equation}

where $A$, $\alpha$ and $\phi$ account for an offset of the anchor of the resonance circle. Taking these effects correctly into account is non-trivial and described for example in [S3,S4]. 

As the transmission response is insensitive to such effects, we first determine the total linewidth by fitting $S_{12}$ to formula S14 and obtain $\kappa_{tot}=3.49$\ MHz. Subsequently we can fit the reflection data as shown in Fig.\@ S3 for $S_{22}$ and determine the coupling rates. From these fits we obtain $\kappa_1=0.51$ \ MHz ($\eta_1=0.147$) and $\kappa_2=2.75$ \ MHz ($\eta_2=0.79$). Using $\kappa_{tot}=\kappa_1+\kappa_2+\kappa_{int}$, we extract the internal loss rate as $\kappa_{int}=230$ \ kHz, corresponding to a $Q_{int}\sim 23.4\times 10^3$. 

\begin{figure}
\includegraphics{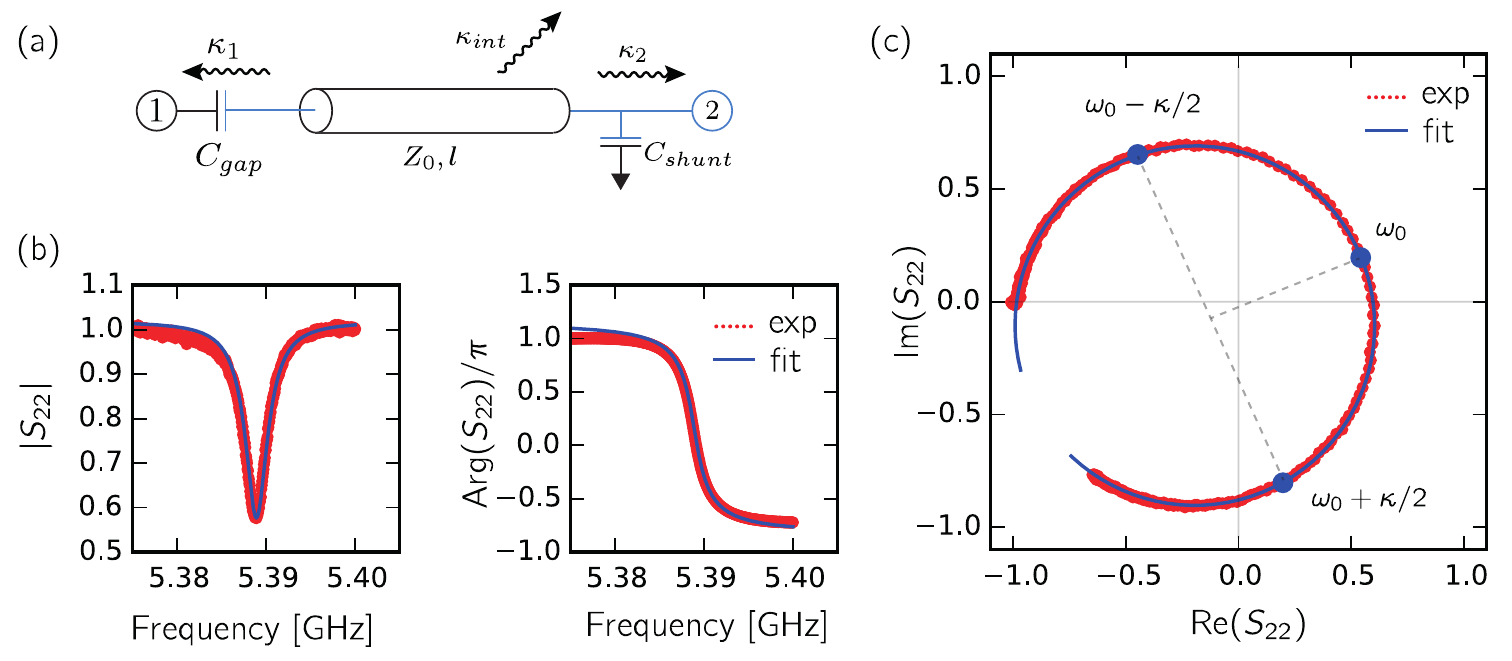}
\caption{\label{fig:wide} Data analysis (a) Schematic of the loss and coupling rates. (b) Reflection of port 2 including the experimental data and fit. (c) Resonance circle plot of the $S_{22}$ response showing data and fit, demonstrating the response is anchored at (-1,0) in the IQ-plane. The fitted resonance frequency and FWHM points are indicated with enlarged blue points.}
\end{figure}



\newpage
\section{Shunt capacitor simulations}
In this section we explore the implications of using the shunt-capacitor coupling architecture on the (potential) performance of a microwave resonator. First we discuss the simulation setup used, and follow with discussing the self-resonances of the shunt capacitor, losses incurred by the dielectric of the capacitor and maximum obtainable Q-factors.
\subsection{Sonnet simulation setup}
Sonnet is simulation software specialized in planar microwave circuits. Both the full cavity and the shunt capacitor are simulated separately, as depicted in Fig.\@ S4. To allow simulations on a $10\times 2.5 \ \mu$m grid we implemented the input coupler as an ideal capacitor of 10 fF. We used a kinetic (sheet) inductance of $L_k=0.78 \ \mbox{pH}/\Box$ for the Molybdenum-Rhenium layers based on earlier measurements for similar films {[}S5{]}. For the SiN brick we used $\epsilon_r=7.4$, and defined $3$ degrees of freedom per lattice point in the $z$-direction. 
\begin{figure}[H]
\includegraphics{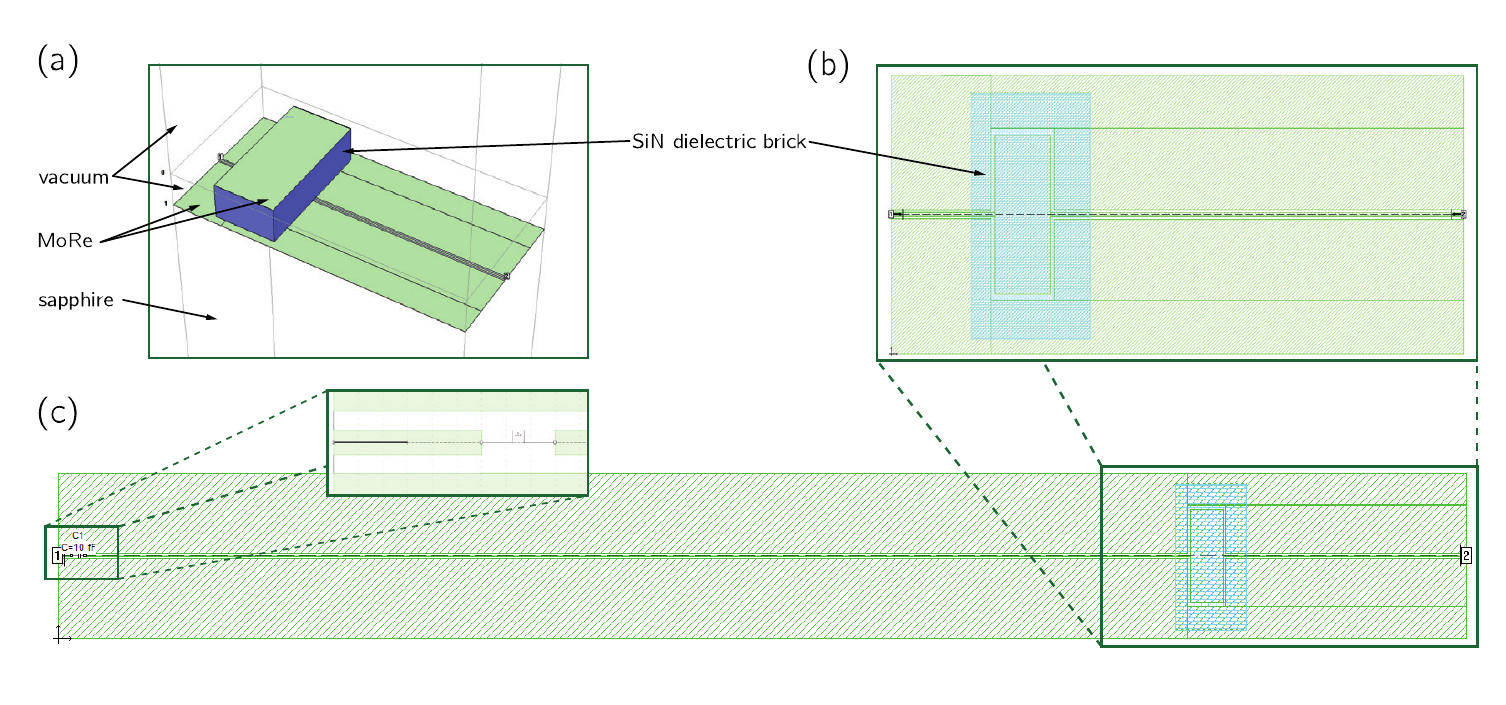}
\caption{\label{fig:wide} Sonnet simulation setup: (a) 3D view of the shunt capacitor showing the substrate, metallization and the SiN dielectric brick defined inside a vacuum layer. (b) Top view of the shunt capacitor simulation. (c) Top view of the simulation configuration where the full bias cavity is simulated. The inset on the left hand side shows the implementation of an ideal input capacitor of $10$\ fF, such that a much lower resolution could be chosen.}
\end{figure}


\subsection{Low frequency response}

Here we simulate the low frequency response of the shunt capacitor as a function of thickness. In this configuration it can be considered as a single-pole low-pass filter, which enables us to determine the capacitance from its cut-off frequency. For applications it can be used to determine the required DC-filtering to protect potential circuits against low frequency noise. Fig.\@ S5a shows that by decreasing the thickness $t$ of the dielectric the transmission for high frequencies is more suppressed, which is expected as $C_{shunt} \propto t^{-1}$ in the parallel plate approximation. More instructive is to consider the transmission on a logarithmic frequency scale, as depicted in Fig.\@ S5b, known as a Bode plot. We observe the slope to be close to 20 dB/octave, characteristic of a single-pole low pass filter. Using a RC-circuit model we can determine the capacitance from the cut-off frequency using $C=1/(R\ 2\pi f_c)$, where $R=25 \ \Omega$ because both $50\ \Omega$ ports are in parallel. For $t=100$\ nm we determine a cut-off frequency of 255 MHz, corresponding to 25 pF of capacitance, close to the 27 pF we expect from a parallel plate approximation. 

\begin{figure}
\includegraphics{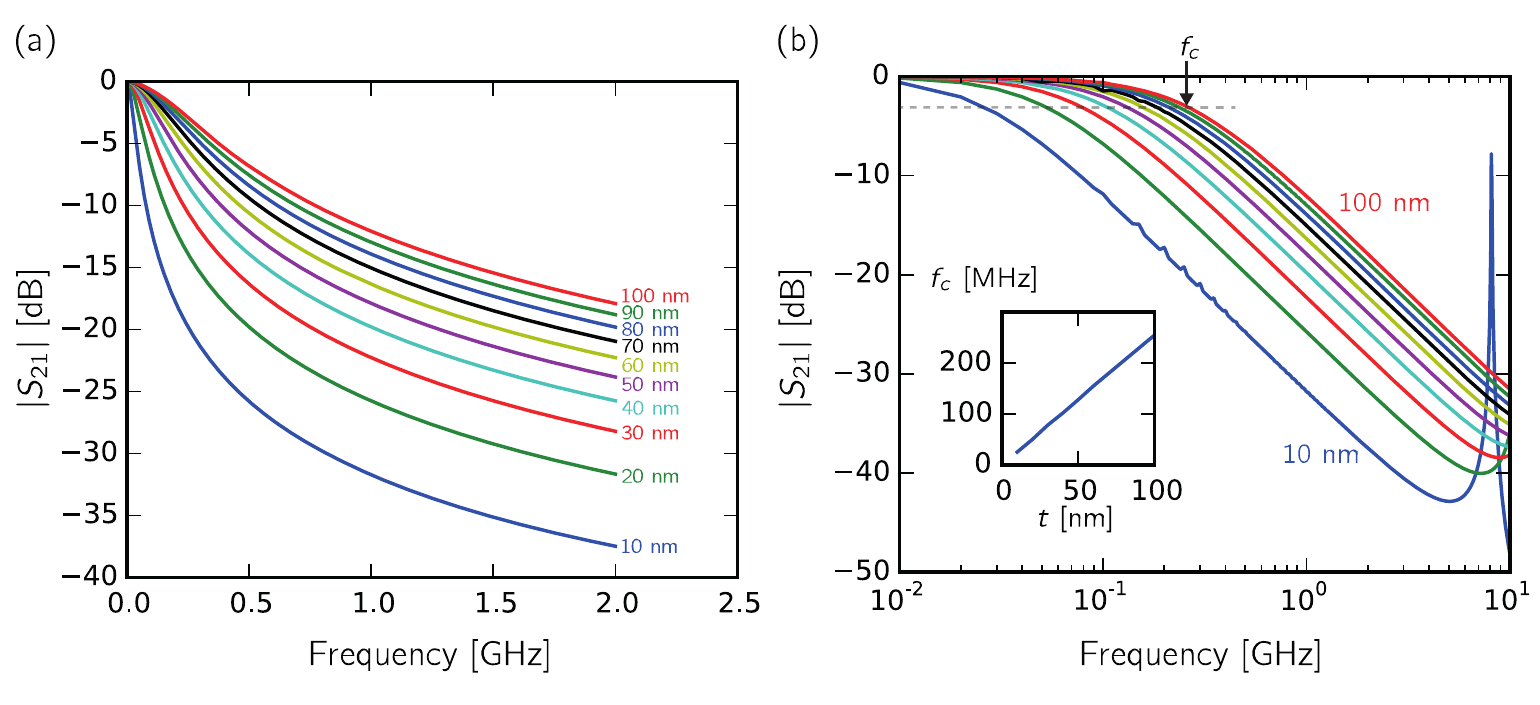}
\caption{\label{fig:wide} Simulated low frequency response of the shunt capacitor, while varying the $\mbox{Si}_3\mbox{N}_4$ dielectric thickness. (a) Transmission response ($|S_{21}|$) as a function of frequency. (b) Transmission ($|S_{21}|$) on a logarithmic frequency scale. We determine the slope as $19.6$ \ dB/octave. For a $t=10$ \  nm we see that the stray inductance causes deviation from this ideal filter model. The inset
 shows the cut-off frequency ($-3$ \ dB-point) as a function of thickness $t$.
 }
\end{figure}

\subsection{Self-resonance frequency}

In the previous section we observed that for a 10  nm dielectric a self-resonance of the shunt capacitor due to stray inductance appeared around $\sim 8$\ GHz. These self-resonances place an upper bound to achievable $Q_c$'s, because the coupling quality-factor increases as the total capacitance increases. To determine the maximum obtainable coupling Q for a $50\ \Omega$ cavity and a shunt capacitor of this geometry, we study these self-resonances as shown in Fig.\@ S6. The first resonance drops from $\sim 25$ GHz for a $100$ nm thickness, to $\sim 8$ GHz for $10$ nm thickness. Thus for circuits requiring coupling Q's in access of $3\cdot10^5$, corresponding $t=10$ nm, one needs to take measures to prevent these resonances to interfere with the circuit, like increasing the impedance of the resonator or introducing structures in the shunt capacitor that prevent box modes (like via's in a pcb).

\begin{SCfigure}[][h!]
  \centering
  \caption{ Simulated microwave transmission, while varying the $\mbox{Si}_3\mbox{N}_4$ dielectric thickness from $100$ nm to $10$ nm in steps of $10$ nm. (a) Transmission response ($|S_{21}|$) of just the shunt capacitor showing self-resonances of the shunt capacitor (dashed grey lines). Each subsequent curve has an offset of $50$ dB. (b) Transmission response for the complete bias cavity showing the first 5 cavity resonances and the self-resonances of the shunt capacitor. Clearly changing the dielectric thickness does not have a noticeable effect on the cavity resonances, but does pull down the self-resonances of the shunt capacitor. }
 \vspace{40pt} \includegraphics
    {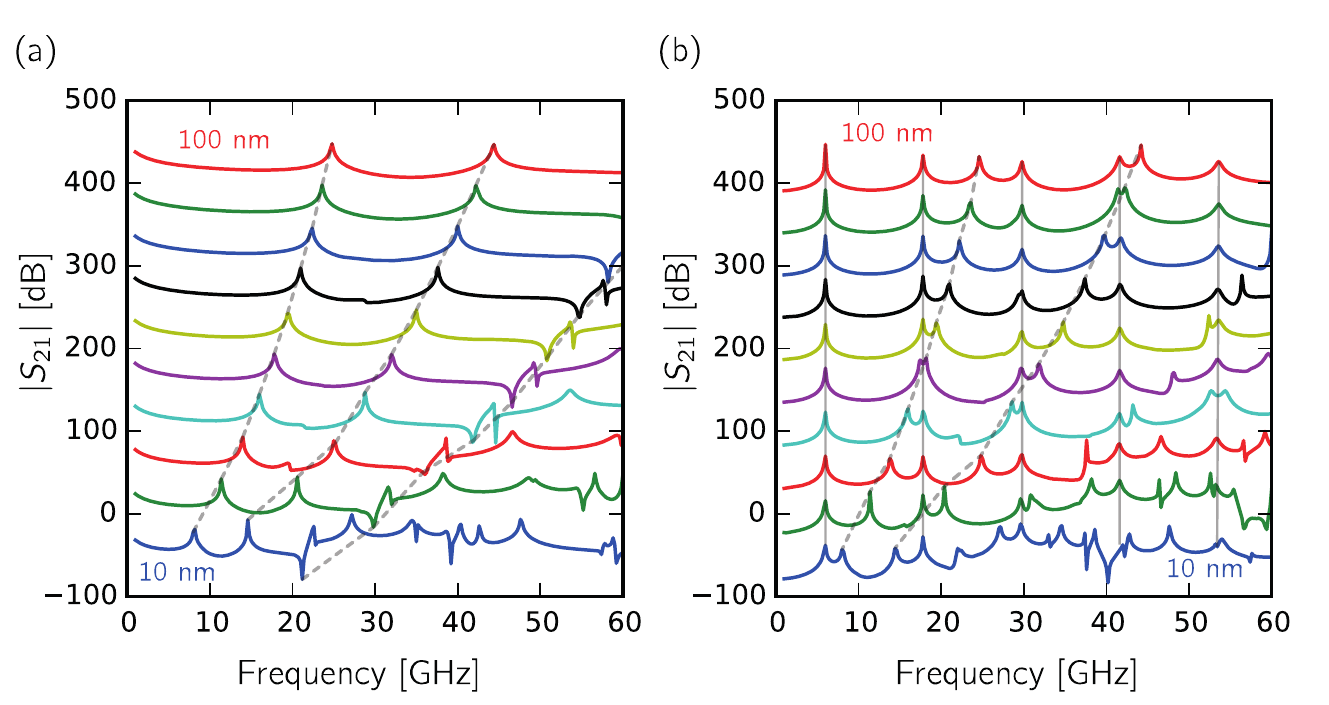}
\end{SCfigure}


\subsection{Dielectric losses}

Deposited dielectrics are well known to be lossy with loss tangents varying from $\tan \delta \sim 10^{-3} - 10^{-5}$. Here we consider this contribution to the losses of a cavity using a shunt capacitor as coupler. The dielectric loss tangent is specified as $\tan \delta = \epsilon_2/\epsilon_1$, with $\tilde{\epsilon}=\epsilon_1+j\epsilon_2$, where $\epsilon_2$ denotes the dissipative part of the dielectric constant. For a lossy parallel plate capacitor, that we showed in previous sections to be a valid approximation, we can use $\tilde{C}=\tilde{\epsilon} A/d=C(1+j \tan \delta)$. As a circuit this can be modelled by including an equivalent series resistor ($R_{esr}$), capturing the losses in the dielectric, where $R_{esr}= \tan \delta /(\omega C)$, as depicted in Fig.\@ S7a. Now we consider a single-port $\lambda/4$ cavity, as displayed in Fig.\@ S7b, and simulate it in QUCs {[}S2{]} for various coupling capacitances and loss tangents. From the simulated scattering data we can extract the coupling rate to port $2$, $\kappa_2$ as a function of dielectric thickness, showed by the dashed line in Fig.\@ S7. As expected $\kappa_2$ 
decreases from about 2 MHz to 20 kHz by decreasing the dielectric thickness. 

\begin{SCfigure}[][h!]
  \centering
  \caption{Lumped element simulation of losses in the shunt capacitor. (a) Incorporating dielectric losses in a capacitor. (b) Lumped element model of single port cavity showing the loss processes. (c) Loss rates as a function of thickness of the capacitor for various dielectric loss tangents. The coupling rate $\kappa_2$ is indicated with the dashed line, the internal losses due to the dielectric losses in the shunt capacitor are shown by the solid lines.
}
 \vspace{40pt} 
 \includegraphics[width=0.8\textwidth]%
    {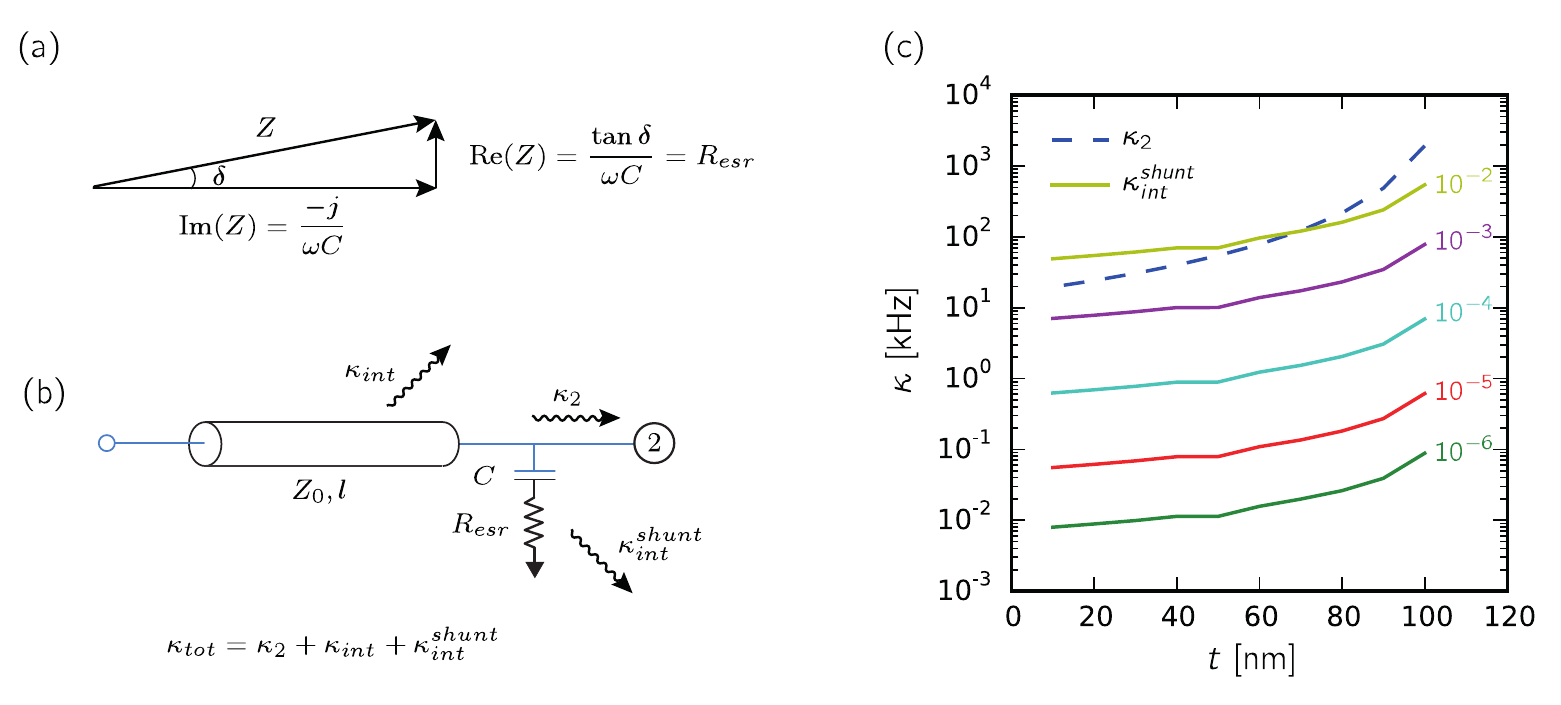}
\end{SCfigure}

\begin{SCfigure}[][h!]
  \centering
  \caption{ Dielectric losses in the shunt capacitor. (a) Figure showing the dielectric losses as a function of loss tangent for various thicknesses. In the inset we see that the losses show a linear relation on a log-log plot indicating a monomial relationship as $\kappa_{int}^{shunt} \propto t \cdot \delta$. (b) Figure showing the maximum obtainable loaded quality factor for different $\delta$ and thicknesses. In this plot other loss factors are ignored and just the coupling rate and dielectric losses in the capacitor are included. 
  }
 \vspace{60pt} 
 \includegraphics[width=0.8\textwidth]%
    {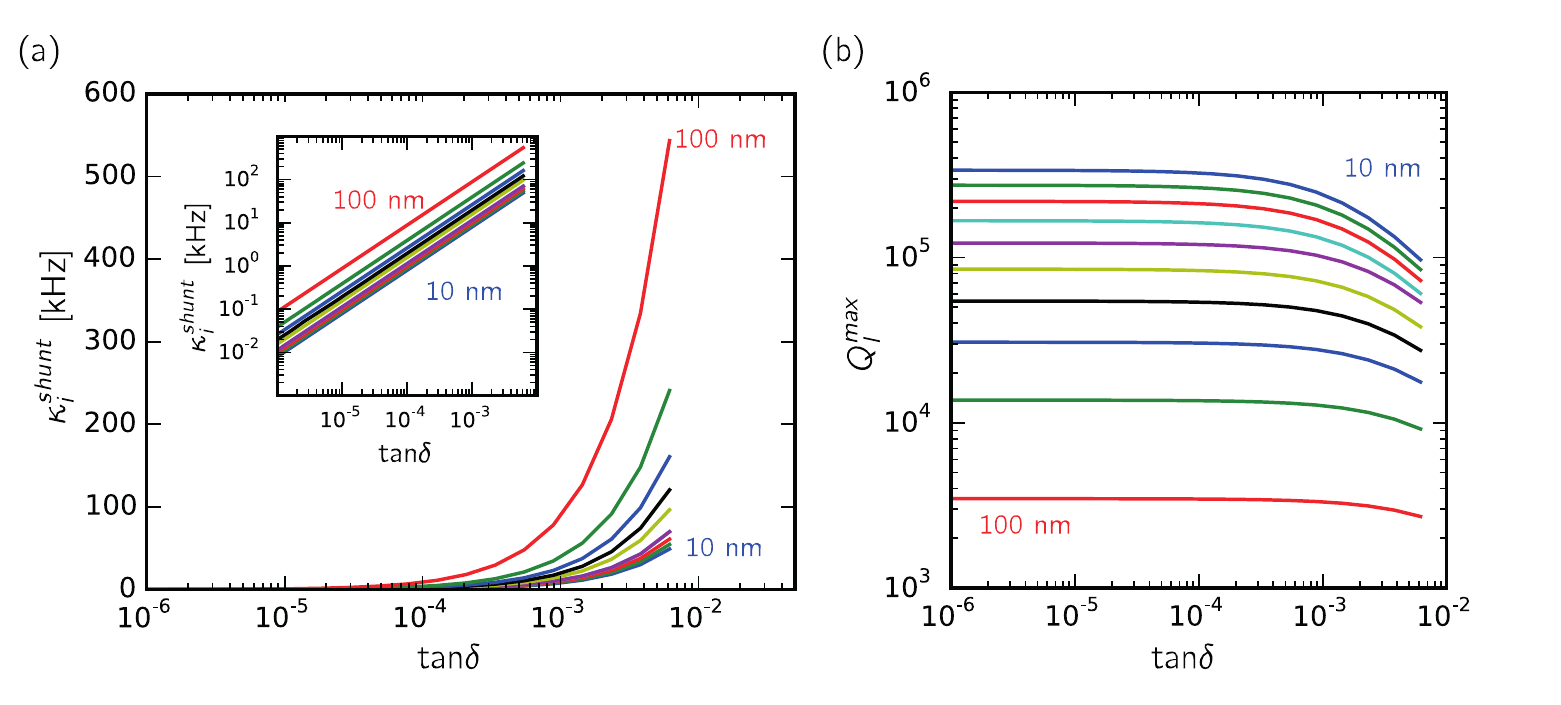}
\end{SCfigure}

%
%

The losses incurred by the lossy dielectric capacitor can be isolated by comparing $\kappa_{int}$ for different $\delta$'s to the lossless case, which is shown in Fig.\@ S7c as the solid lines. These incurred losses decrease as the coupling capacitance increases, because the voltage across the capacitor plates decreases and hence the dielectric losses decrease as well. Assuming a loss tangent of around $\delta \sim 10^{-3} - 10^{-4}$ we estimate these losses in our experiment to be on the order of $\kappa^{shunt}_{int} \sim 8-80$ \ kHz. Finally from this figure we see that for thin dielectrics and realistic $\delta$'s the incurred losses can be minimized to sub-kHz level.

Finally we can use this data to determine what the maximum possible loaded quality factor for this resonator architecture is, shown in Fig.\@ S8. By including both the coupling rate and dielectric losses of the capacitor, we see that the loaded quality factor is limited to $Q_l \lesssim 3 \times 10^{5}$ for a dielectric thickness of 10 nm. Such a loaded quality factor corresponds to linewidth's in the range of $\kappa_{tot} \gtrsim 20$\ kHz, sufficiently narrow for most applications. Beyond this regime a different geometry or a higher impedance resonator is required.


\ \\ \ \\
{[}S1{]} D. M. Pozar, \textit{Microwave engineering} (John Wiley \& Sons, 2009) \\
{[}S2{]} M. Brinson and S. Jahn, International Journal of Numerical Modelling: Electronic Networks, Device and Fields \\ \hspace*{13pt}  \ \textbf{22}, 297 (2009) \\
{[}S3{]} M. Khalil, M. Stoutimore, F. Wellstood, and K. Osborn, Journal of Applied Physics 111, 054510 (2012).\\
{[}S4{]} S. Probst, F. Song, P. Bushev, A. Ustinov, and M. Weides, Review of Scientific Instruments \textbf{86}, 024706 (2015). \\
{[}S5{]} V. Singh, B. H. Schneider, S. J. Bosman, E. P. Merkx, and G. A. Steele, Applied Physics Letterrs \textbf{105}, 222601 \\ \hspace*{15pt} (2014). \\


\end{document}